\documentstyle[12pt]{article}
\begin{document}
\baselineskip .4in
\pagestyle{empty}
\begin{center}
{\Large{\bf Critical behaviour in $La_{0.5}Sr_{0.5}CoO_{3}$}}\\
\end{center}
\vskip 2.0cm
\begin{center}
S. Mukherjee, P. Raychaudhuri, A. K. Nigam\\
Tata Institute of Fundamental Research,\\
Homi Bhabha Road, Colaba, Mumbai - 400 005, India.\\
\end{center}
\vskip 1.0cm

\newpage
\pagestyle{empty}
\begin{center}
Abstract
\end{center}
We have studied the critical behaviour in $La_{0.5}Sr_{0.5}CoO_{3}$ near 
the paramagnetic-ferromagnetic transition temperature. We have analysed 
our dc magnetisation data near the transition temperature with the help of 
modified Arrott plots, Kouvel-Fisher method. We have determined 
the critical temperature $T_{c}$ and the critical exponents, $\beta$ and 
$\gamma$. With these values of $T_c$, $\beta$ and $\gamma$, we plot 
$M/(1-T/T_c)^{\beta}$ vs $H/(1-T/T_c)^{\gamma}$. All the data collapse on 
one of the two curves. This suggests that the data below and above $T_c$ 
obeys scaling, following a single equation of state.
\newpage
\pagestyle{empty}
\begin{center}
{\bf I. INTRODUCTION }
\end{center}
\vskip 1.0cm

The parent compound $LaCoO_3$ has been studied extensively$.^{1-4}$ This 
is a semiconductor with a conductivity gap of almost 0.6 eV and shows a 
decrease in susceptibility below 100 K with a broad maximum and a gradual 
reduction obeying Curie-Weiss law at higher temperatures. The magnetic 
properties of the system are explained by a progressive conversion of 
low-spin $Co^{III}$ into high-spin $Co^{3+}$. In the temperature range 
$100 K < T < 350 K$, the ratio of the high-spin to low-spin Co reaches 
50:50 with short range ordering of low-spin and high-spin Co ions. Above 
600 K it behaves metallic. Substitution of $Sr^{2+}$ for $La^{3+}$ causes 
a remarkable change in the system. Due to Sr doping the material segregates 
into hole-rich, metallic ferromagnetic regions and a hole poor matrix like 
$LaCoO_3$. The Co ions in the ferromagnetic phase are in intermediate-spin 
configurations$.^5$ In $La_{1-x}Sr_{x}CoO_{3}$, for $x < 0.2$, the hole rich regions are isolated from each 
other and show superparamagnetic behaviour below $T_c \sim 240 K.^ {5,6}$ These 
clusters freeze at a lower temperature. For $x > 0.2$, the onset of ferromagnetic 
transition is observed $.^{5,7,8}$ Metallic ferromagnetism has been suggested for the 
range $0.30 \leq x \leq 0.50.^ 5$ However, the hole-poor matrix 
interpenetrating the ferromagnetic regions persists to x = 0.5$.^{5,6,9}$ 

  In order to fully understand the nature of the ferromagnetic transition 
we carried out  the study of the critical exponents in detail associated with 
the transition. For our study we have chosen the extreme ferromagnetic limit 
$La_{0.5}Sr_{0.5}CoO_3$. Experimental studies of critical phenomena have 
been previously made  on manganese oxides $.^ {10 - 12}$ Recently similar study has been 
done in $La_{1-x}Sr_{x}MnO_3$ system $.^{13,14}$ 

\vskip 1.0cm
\begin{center}
{\bf II. EXPERIMENTAL}
\end{center}
\vskip 1.0cm

The sample, $La_{0.5}Sr_{0.5}CoO_{3}$ was prepared by a solid state reaction 
method starting with preheated $La_2O_3$, CoO and $SrCO_3$. The appropriate 
mixture was ground and calcined at $1000^{o}C$ for 1 day. The mixture was 
then ground again and heated at $1100^{o}C$ in air for 2 days with 
intermediate grindings.  It was then pelletized and fired in air at
$1300^oC$ for 1 day. The phase purity was checked with X rays and the sample 
was found to be of single phase and the diffraction pattern compared well 
with the reported data.

The magnetisation measurements were performed using SQUID magnetometer (
Quantum Design). The data were collected at 2 K intervals over the 
temperature range from 202 K to 270 K, in fields from 100 Oe to 55 kOe. 
The maximum deviation in the temperature was $\pm  0.02$ K at each measuring temperature.

\vskip 1.0cm
\begin{center}
{\bf III. RESULTS AND DISCUSSION}
\end{center}
\vskip 1.0cm

The second-order magnetic phase transition near the Curie point is 
characterised by a set of critical exponents, $\beta$(associated with 
the spontaneous magnetisation), $\gamma$ (associated with the initial 
susceptibility) and $\delta$ (related to the critical magnetisation 
isotherm). They are defined $as ^ {14}$ 
\begin{equation}
M_s (T) = M_0(-\epsilon)^{\beta}, \epsilon < 0,
\end {equation}
\begin{equation}
 \chi_0^{-1}(T) = (h_{0}/M_0)\epsilon^{\gamma}, \epsilon > 0,
\end{equation}
\begin{equation}
M = A_0(H)^{1/\delta}, \epsilon = 0.
\end{equation}
where $\epsilon = (T - T_c)/T_c$, $T_c$ is the Curie temperature and 
$M_0$, $h_0/M_0$ and $A_0$ are the critical amplitudes. 
Our aim is to determine the critical exponents and the critical temperature 
from the magnetisation data as a function of the field at different 
temperatures. 

Figure 1 shows the zero field cooled (ZFC) and field cooled (FC) 
magnetisation data as a function of temperature in a field of 100 Oe. 
A sharp rise in the ZFC magnetisation below 240 K is considered as the 
signature of ferromagnetic ordering. Below the Curie temperature, 
FC magnetisation value is different from the ZFC value. Figure 2 shows 
hysteresis loop at 5 K. The loop is similar to that of a ferromagnet with 
small coercive force and remanence. 

Figure 3 and figure 4 show the magnetisation data as a function of the 
magnetic field at different temperatures. Figure 5 shows the $M^2$ vs 
$H/M$ plot or the Arrot plot. According to the mean field theory near 
$T_c$, $M^2$ vs $H/M$  at various temperatures should show a series of 
parallel lines. The line at T = $T_c$ should pass through the origin.
In our case the curves in the Arrot plot are not linear. This suggests 
that the mean field theory is not valid. Then we tried to analyse our 
data according to the modified Arrot plot method.  This method is based 
on the Arrot-Noakes equation of state $.^ {15}$ According to this method, the plot 
of $M^{1/\beta}$ and $(H/M)^{1/\gamma}$ at several temperatures close to 
$T_c$ should give a series of parallel straight lines. We adopted this 
method and obtained almost parallel straight lines with the values of 
$\beta$ and $\gamma$ equal to  0.365 and 1.336 respectively. Figure 6 shows the modified 
Arrot plot in our case. This plot shows that the isotherms are almost 
parallel straight lines in the high-field region.  The high field straight 
line portions of the isotherms can be linearly extrapolated to obtain the 
intercepts on the $M^{1/\beta}$ and $(H/M)^{1/\gamma}$ axes. From these 
intercepts the spontaneous magnetisation $M_s(T)$ and the inverse 
susceptibility $\chi_0^{-1}(T)$ can be computed. The temperature variation 
of $M_s(T)$ and the same for $\chi_0^{-1}(T)$, obtained from figure 6 
are shown in figure 7 and figure 8. The continuous curves in figures 
7 and 8 show the power-law fit obtained from equation (1) and (3) 
respectively. Then we apply Kouvel-Fisher method$^{16}$ to obtain $T_c$, $\beta$
and $\gamma$. The Kouvel-Fisher method suggests that the quantities 
$M_s(T)(dM_s(T)/dT)^{-1}$ and $\chi_{0}^{-1}(T)(d\chi_0^{-1}(T) /dT)^{-1}$
plotted against temperature, give straight lines with slopes (1/$\beta$) 
and (1/$\gamma$) respectively, and the intercepts on the T-axes are equal 
to $T_c/\beta$ and 
$T_c/\gamma$ respectively. From figures 7 and 8 we have computed 
$M_s(T)(dM_s(T)/dT)^{-1}$ and $\chi_{0}^{-1}(T)(d\chi_0^{-1}(T) /dT)^{-1}$. 
Figure 9 and figure 10 show respectively the plots of 
$M_s(T)(dM_s(T)/dT)^{-1}$ and $\chi_{0}^{-1}(T)(d\chi_0^{-1}(T) /dT)^{-1}$
vs temperature. The linear fit to the plot shown in figure 9 gives the 
value of $\beta$ as 0.321 $\pm$ 0.002 and $T_c$ as 222.82 K. The linear fit 
to the 
plot shown in figure 10 gives the value of $\gamma$ as 1.351 $\pm$ 0.009 
and $T_c$ as 223.18 K. Figure 11 shows M vs H plot on a log scale at few 
temperatures close to $T_C$. The straight line shows the fit for the 
interpolated data at $T_c$ = 223 K. This gives the value of $\delta$ as 
4.39 $\pm$ 0.02.

  Next we compare our data with the prediction of the scaling theory$^{17}$
\begin{equation}
M/|\epsilon|^{\beta} = f_{\pm}(H/|\epsilon|^{(\beta + \gamma)})
\end{equation}
where (+) and (-) signs are for above and below $T_c$ respectively.
This relation further predicts that $M/|\epsilon|^{\beta}$ plotted 
as a function of $H/|\epsilon|^{\beta + \gamma}$ give two different 
curves, one for temperatures below $T_c$ and the other for temperatures 
above $T_c$. Taking the values of $\beta$, $\gamma$ and $T_c$ obtained 
from our analysis, the scaled data are plotted in figure 12. All the 
points fall on two curves, one for $T < T_c$ and the other for $T > T_c$. 
This suggests that the value of the exponents and $T_c$ are accurate 
enough.

 We started analysing our data with the help of modified Arrot plot.
Finally we got the exponent values as $\beta = 0.321 \pm 0.002$, 
$\gamma = 1.351 \pm 0.009$, $\delta = 4.39 \pm 0.02$. Our experimental 
value of $\beta$ is very close to 3D  Ising value. However, the value 
of $\gamma$ and $\delta$ do not match very accurately.  
The value of the critical exponents depend on the 
range of the exchange interaction J(r). According to Fisher et al$.^ {18}$  
J(r) varies as $1/r^{d+\sigma}$, where d gives the dimension of the system 
and $\sigma$ gives a measure of the range of the interaction. If $\sigma$ 
is greater than 2, then the Heisenberg exponents ($\beta$ = 0.365, 
$\gamma$ = 1.386 and $\delta$ = 4.8) are valid. The mean field exponents 
($\beta$ = 0.5, $\gamma$ = 1 and $\delta$ = 3.0) are valid for  $\sigma $
less than half. For $1/2 < \sigma <2$, the exponents belong to different 
universality class which depends upon $\sigma$. $La_{1-x}Sr_{x}CoO_{3}$ is not a very simple ferromagnetic system. It shows a large difference 
between the ZFC and FC magnetisation below $T_c$. The hole-poor matrix is also present in $La_{0.5}Sr_{0.5}CoO_{3}.^{5}$ Hence one cannot expect the exponents to belong to any universality class.

\vskip 1.0cm
\begin{center}
{\bf IV. CONCLUSION}
\end{center}
\vskip 1.0cm

We have studied the the critical behaviour of $La_{0.5}Sr_{0.5}CoO_{3}$ 
polycrystalline sample from dc magnetisation measurement near $T_c$. 
We have determined the values of $T_c$, $\beta$, $\gamma$, $\delta$. 
The value of $\beta$ is very close to 3D Ising value. However, the values 
of $\gamma$ and $\delta$ do not completely agree with any universality 
class.

\vskip 1.0cm
\begin{center}
{\bf V. ACKNOWLEDGMENT}
\end{center}

We thank K.V. Gopalakrishnan for his sincere help in the experiment.

\newpage
\vskip 1.0cm
\begin{center}
{\bf REFERENCES}
\end{center}
\vskip 1.0cm

\begin{enumerate}

\item J.B. Goodenough, J. Phys. Chem. Solids {\bf 6}, 287 (1958).
\item G.H. Jonker, J. Appl. Phys. {\bf 37}, 1424 (1966). 
\item P.M. Raccah and J.B. Goodenough, Phys. Rev. {\bf 155}, 932 (1967).
\item M.A. Senaris-Rodriguez and J.B. Goodenough, J. Solid State Chem. {\bf 116}, 224 (1995).
\item M.A. Senaris-Rodriguez and J.B. Goodenough, J. Solid State Chem. {\bf 118}, 323 (1995).
\item M. Itoh, I. Natori, S. Kuboto and K. Motoya, J. Phys. Soc Jpn {\bf 63}, 1486 (1994).
\item R. Mahendiran and A.K. Raychoudhuri, Phys. Rev. B {\bf 54}, 16044 (1996).
\item V.G. Sathe, A.V. Pimpale, V. Siruguri and S.K. Paranjpe, J. Phys. CM {\bf 8}, 3889 (1996).
\item S. Mukherjee, R. Ranganathan, P.S. Anilkumar and P.A. Joy, Phys. Rev. B {\bf 54} 9267 (1996).
\item S.E. Lofland, V. Ray, P.H. Kim, S.M. Bhagat, M.A. Manheimer, S.D. Tyagi, Phys. Rev. B {\bf 55} 2749 (1997).
\item M.C. Martin, G. Shirane, Y. Endoh, K. Hirota, Y. Moritomo, Y. Tokura Phys. Rev. B {\bf 53}, 14285 (1996).
\item J. Lynn et al., Phys. Rev. Lett. {\bf 76} 4046 (1996).
\item K. Ghosh, C.J. Lobb, R.L. Greene, S.G. Karabashev, D.A. Shulyatev, A.A. Arsenov and Y. Mukovskii.
\item
Ch. V. Mohan, M. Seeger, H. Kronmuller, P. Murugaraj and J. Maier, J. Magn Magn Mater. {\bf 183}, 348 (1998).
\item A. Arrot and J.E. Noakes, Phys. Rev. Lett {\bf 19}, 786 (1967).
\item J.S. Kouvel and M.E. Fisher, Phys. Rev. A {\bf 136}, 1626 (1964).
\item H. Eugene Stanley, Introduction to phase transitions and Critical phenomena, Oxford University Press, New York, 1971.
\item M.E. Fisher, S.K. Ma, B.G. Nickel, Phys. Rev. Lett {\bf 29}, 917 (1972).

\end{enumerate}

\newpage
\vskip 1.0cm
\begin{center}
{\bf FIGURE CAPTIONS}
\end{center}
\vskip 1.0cm

{\bf figure 1}:- Magnetisation as a function of temperature under zero field cooled (ZFC) and field cooled (FC) conditions. 
\vskip 0.3cm

{\bf figure 2}:- Hysteresis loop drawn at 5 K.\vskip 0.3 cm

{\bf figure 3}:- Magnetisation as a function of magnetic field at several temperatures in the range 202 K - 232 K. 
\vskip 0.3 cm

{\bf figure 4}:-Magnetisation as a function of the magnetic field at several temperatures in the range 234 K - 270 K. 
\vskip 0.3 cm

{\bf figure 5}:- Isotherms of $M^2$ vs H/M.
\vskip 0.3 cm

{\bf figure 6}:- Modified Arrot plot isotherms.
\vskip 0.3 cm

{\bf figure 7}:- The temperature variation of the spontaneous magnetisation along with the fit obtained with the help of the power law.
\vskip 0.3 cm

{\bf figure 8}:- The temperature variation of the inverse initial susceptibility along with the fit obtained with the help of the power law.
\vskip 0.3 cm

{\bf figure 9}:- Kouvel-Fisher plot for the spontaneous magnetisation
\vskip 0.3 cm

{\bf figure 10}:- Kouvel-Fisher plot for the inverse initial susceptibility.
\vskip 0.3 cm

{\bf figure 11}:- M vs H on a log scale at several temperatures close to $T_c$.
\vskip 0.3 cm

{\bf figure 12}:- The scaling plot on a log scale. 
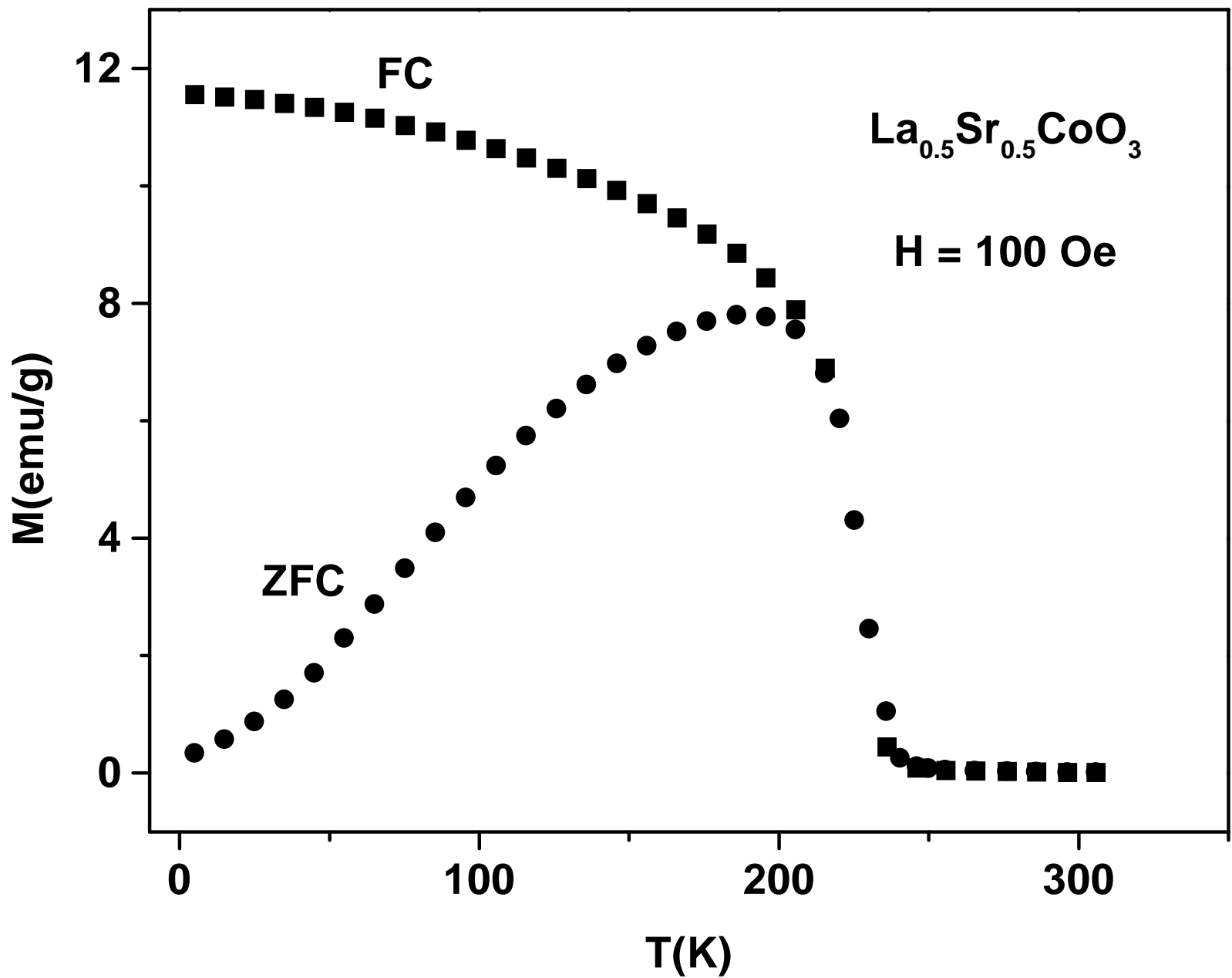
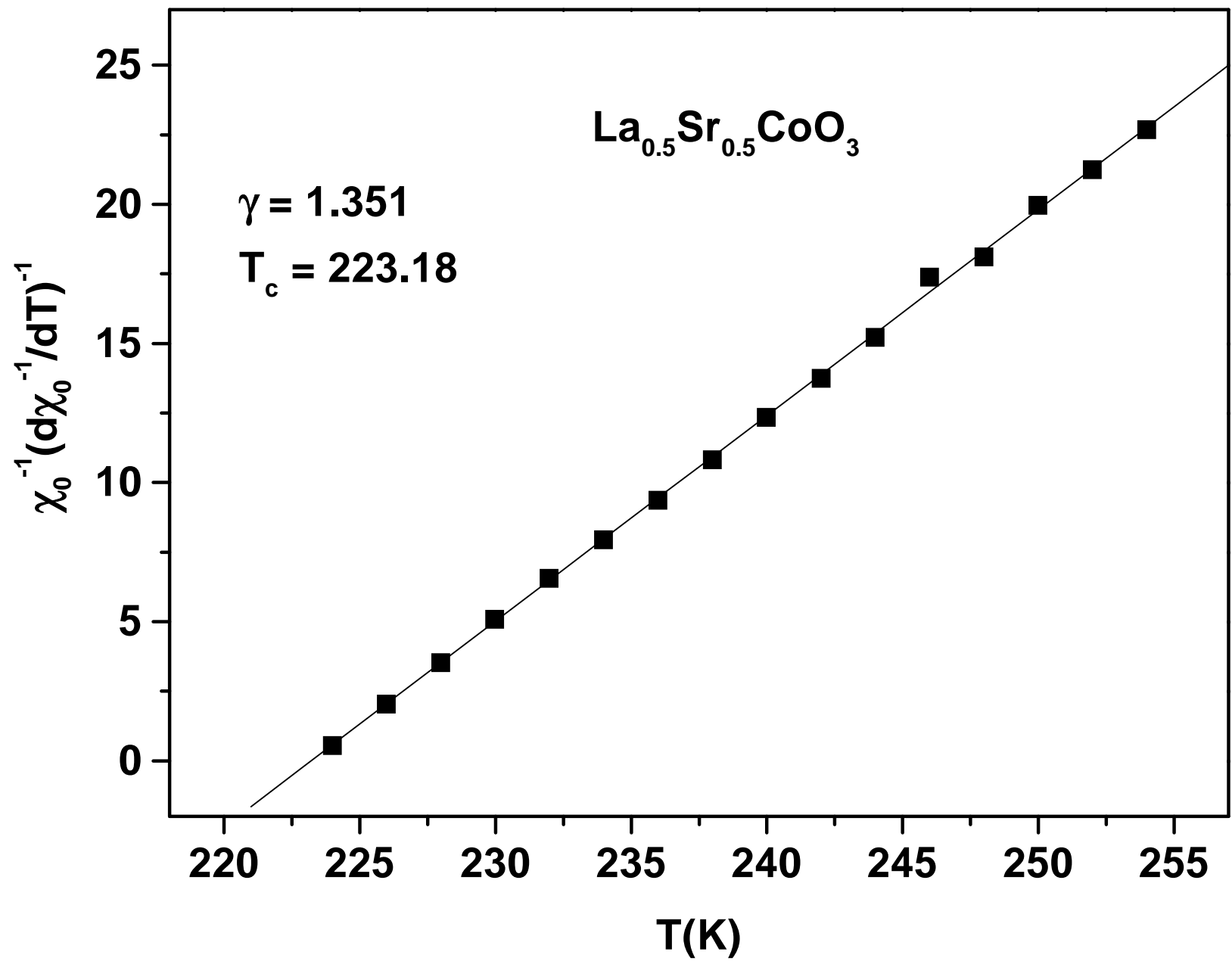
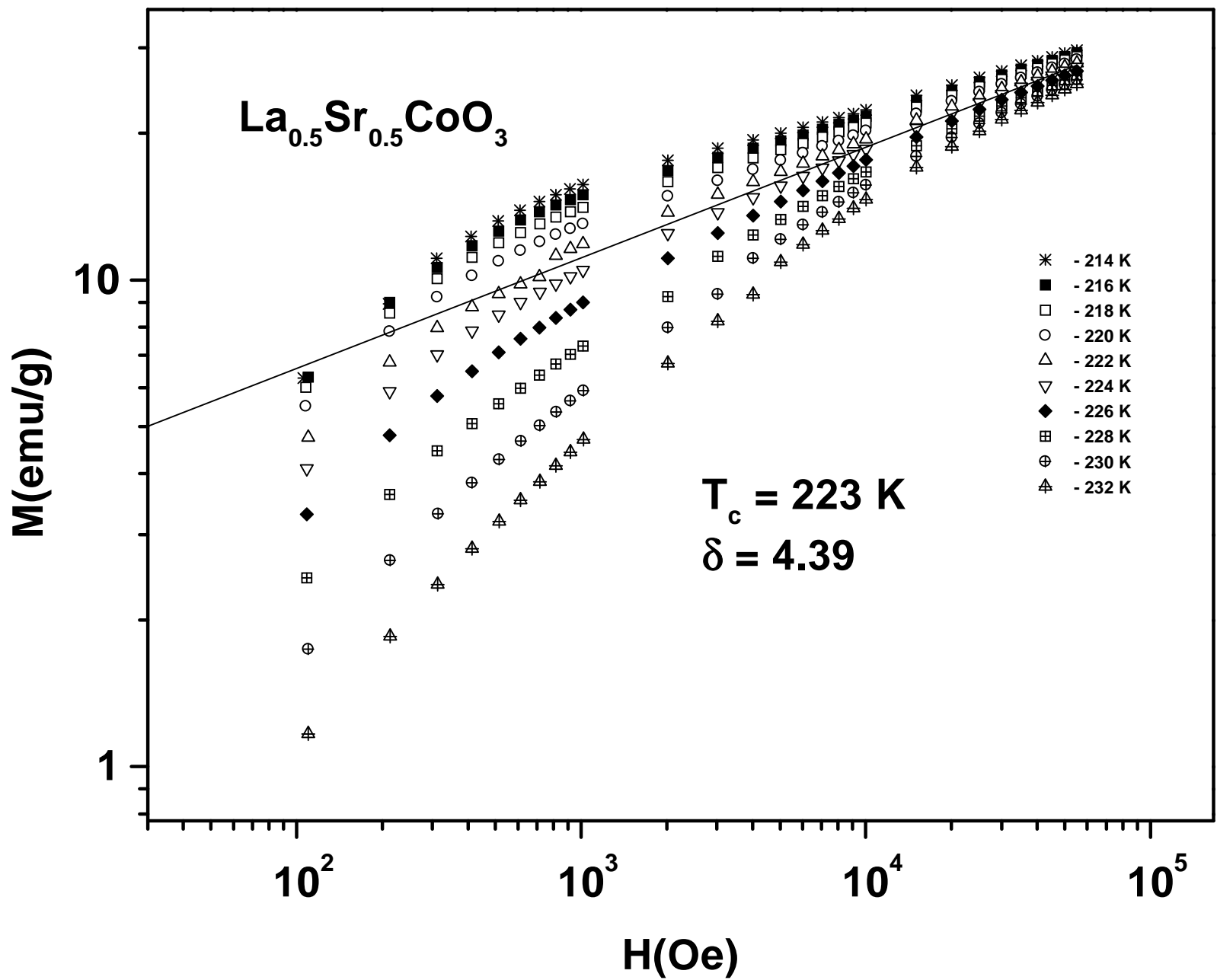
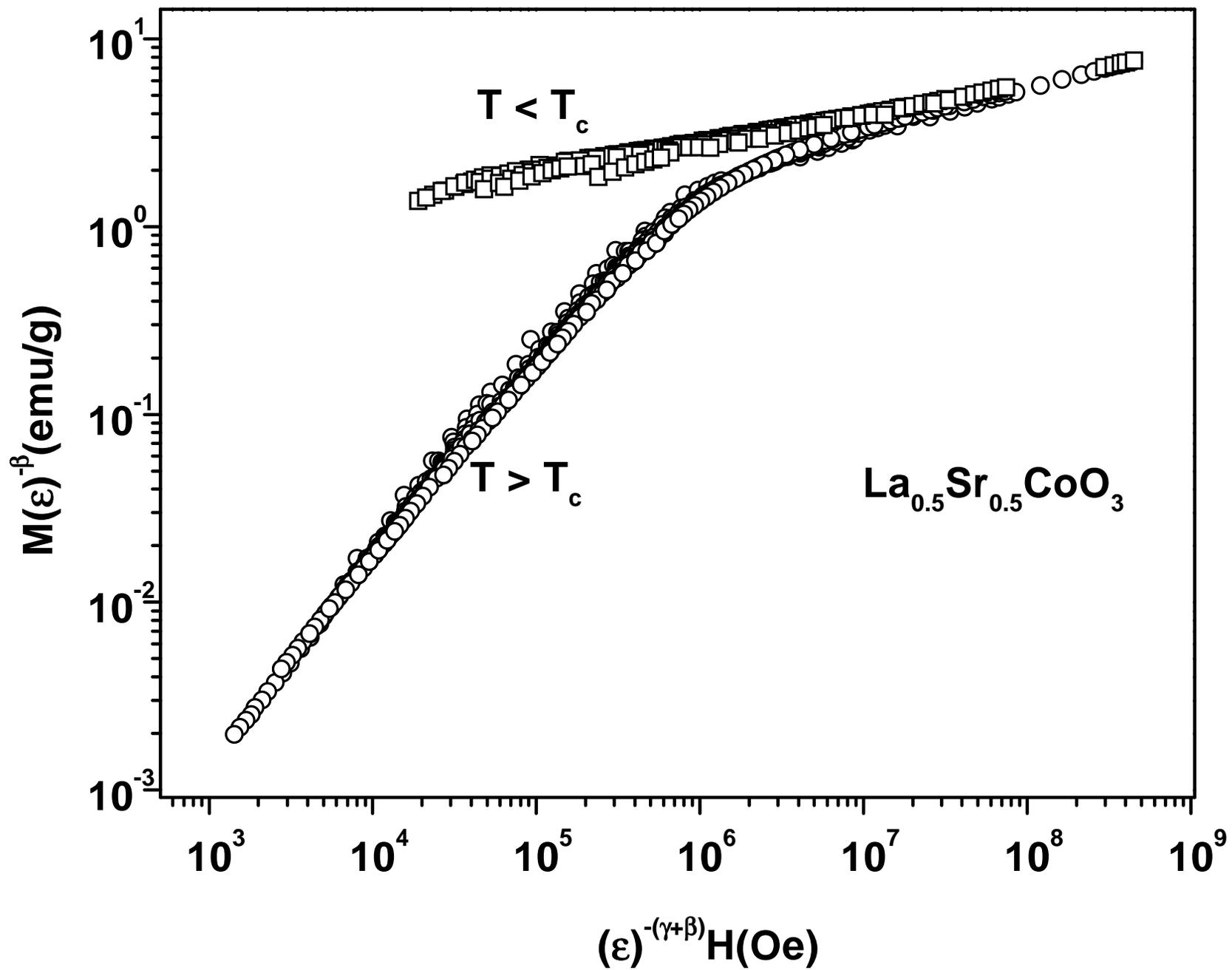
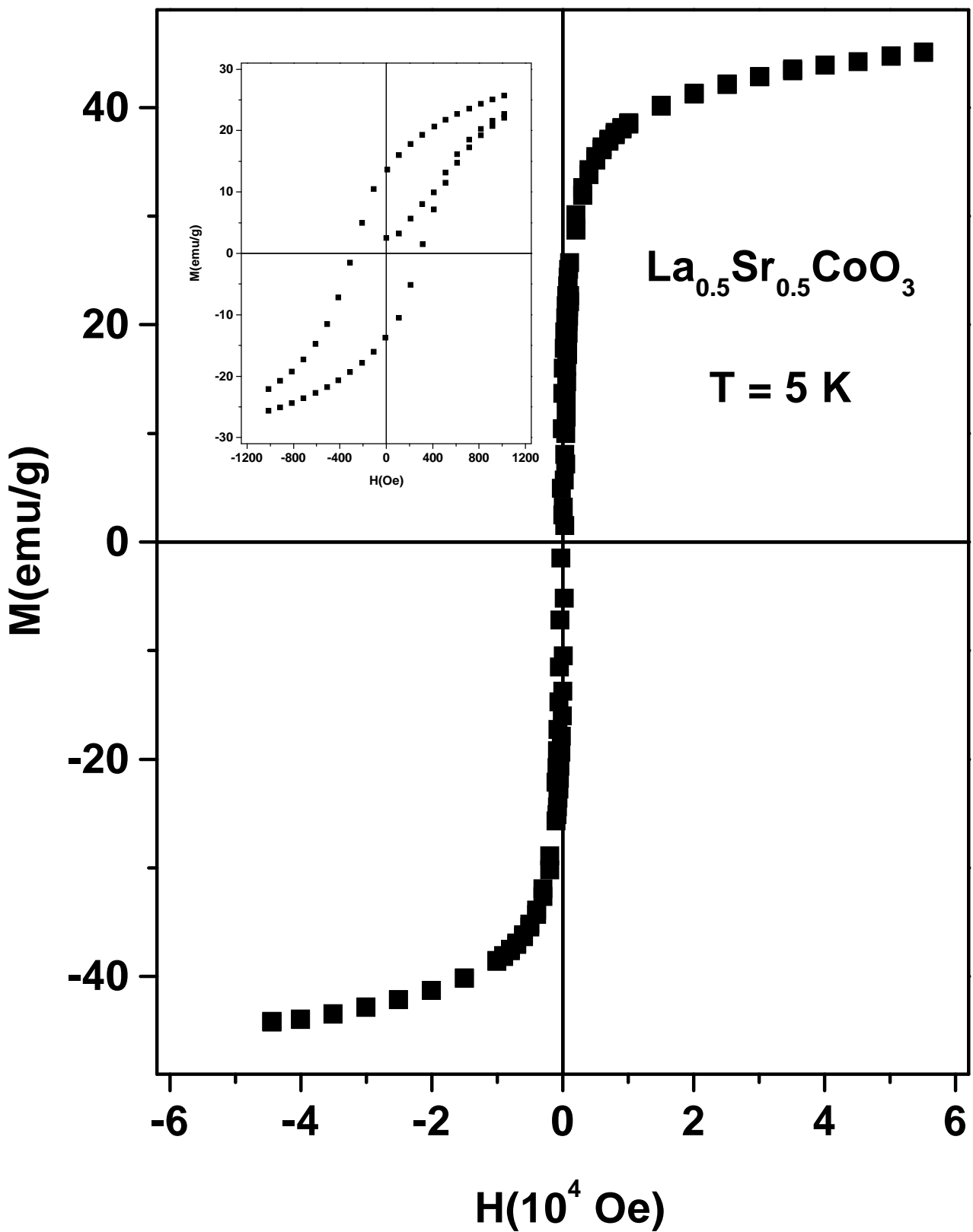
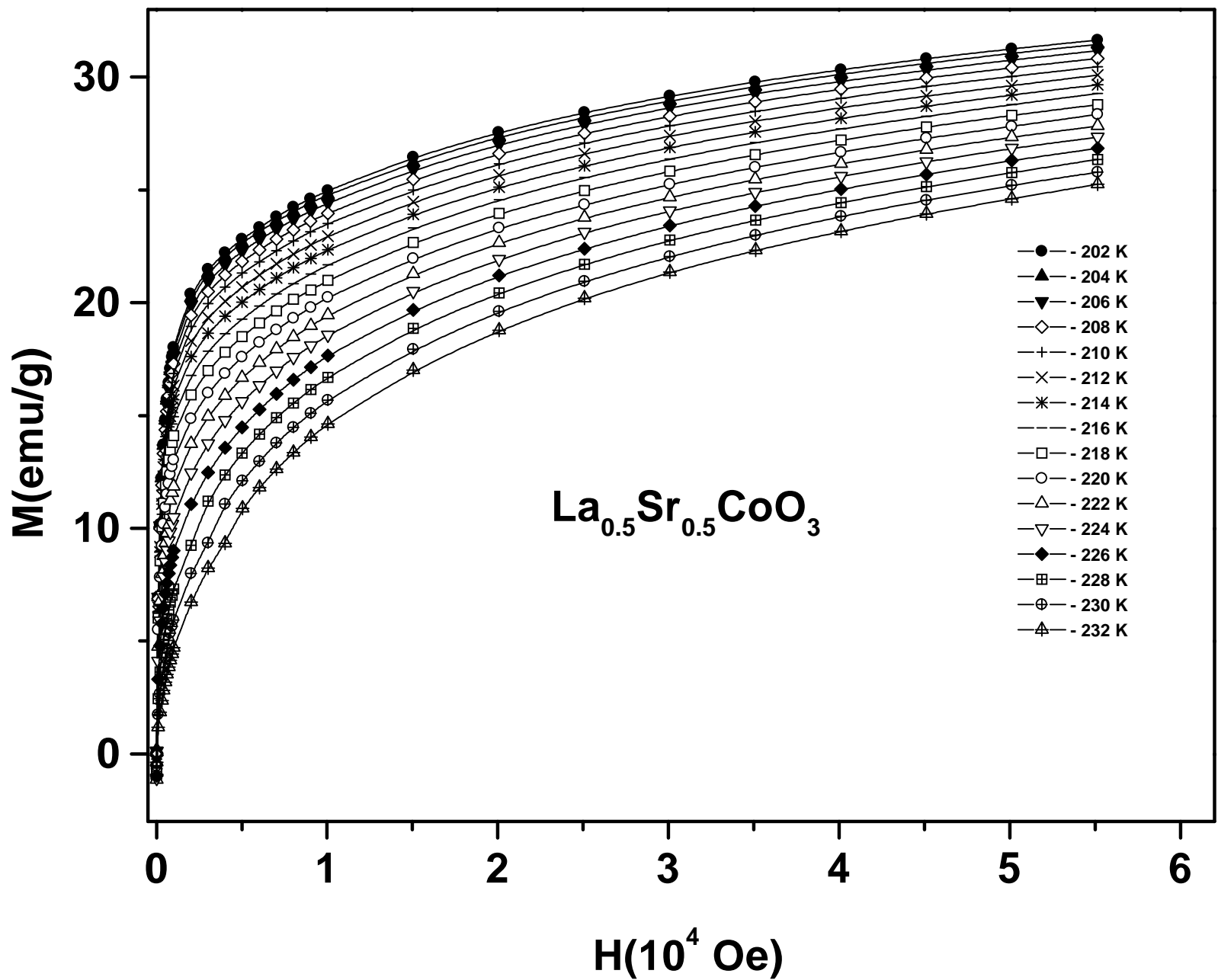
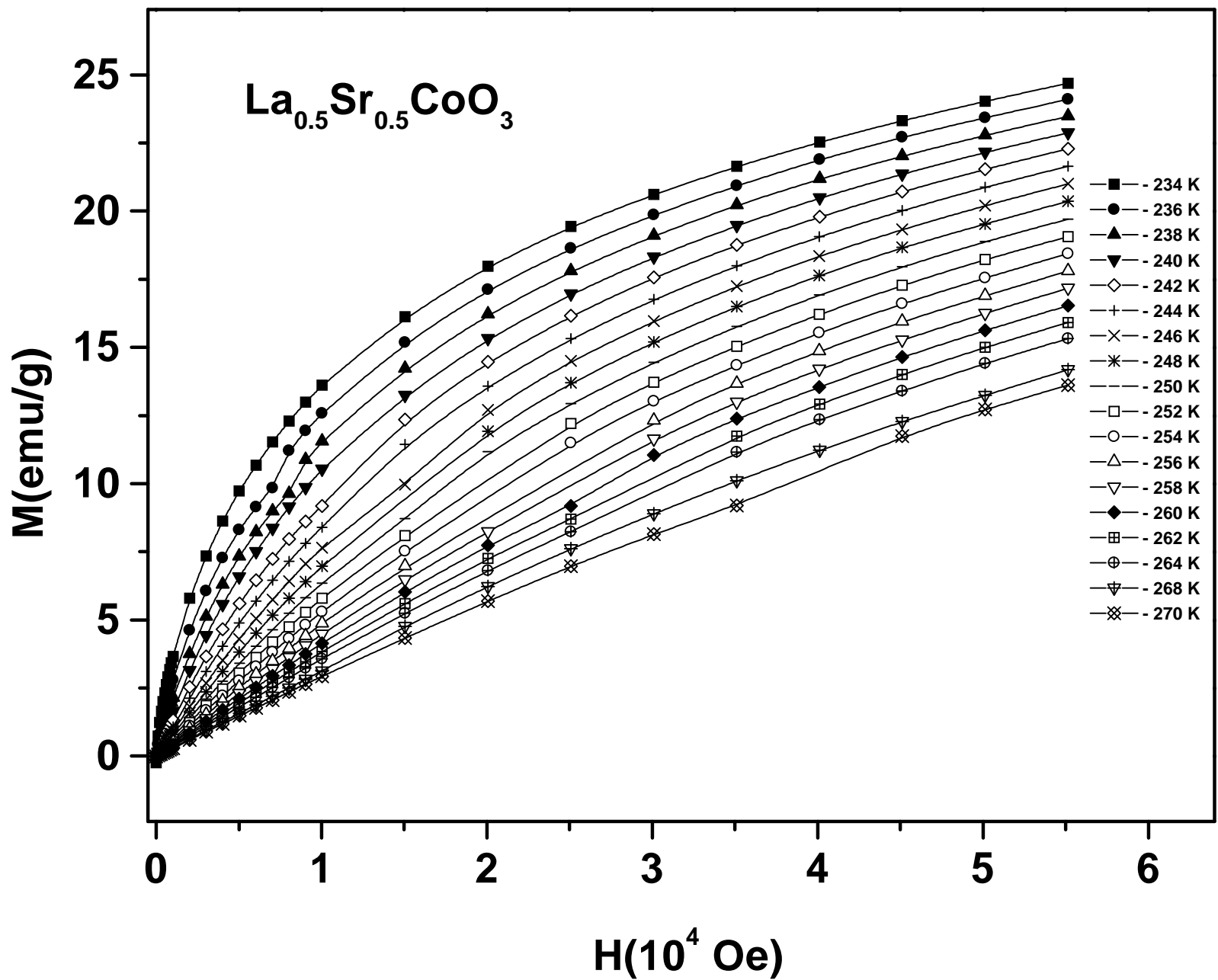
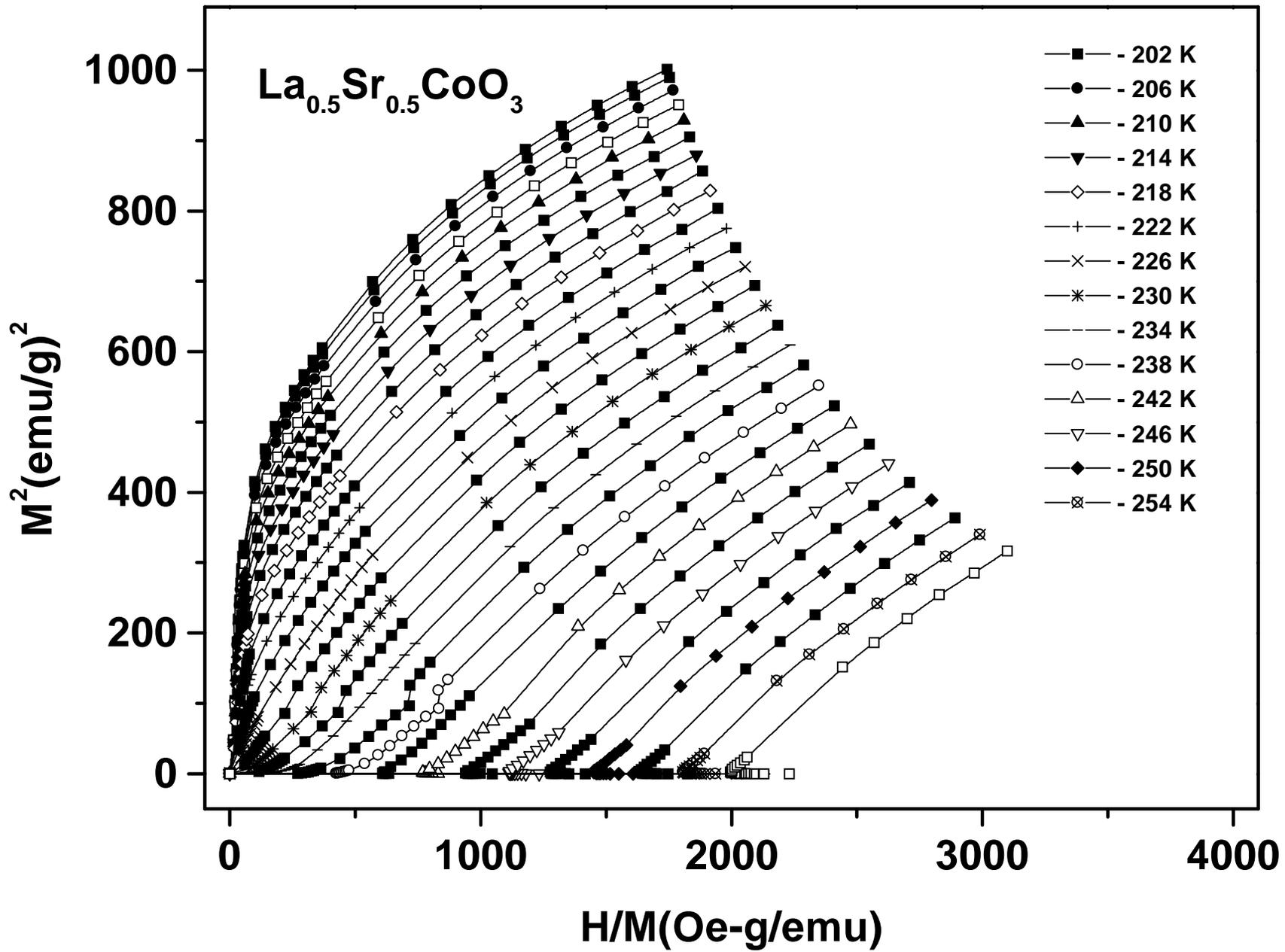
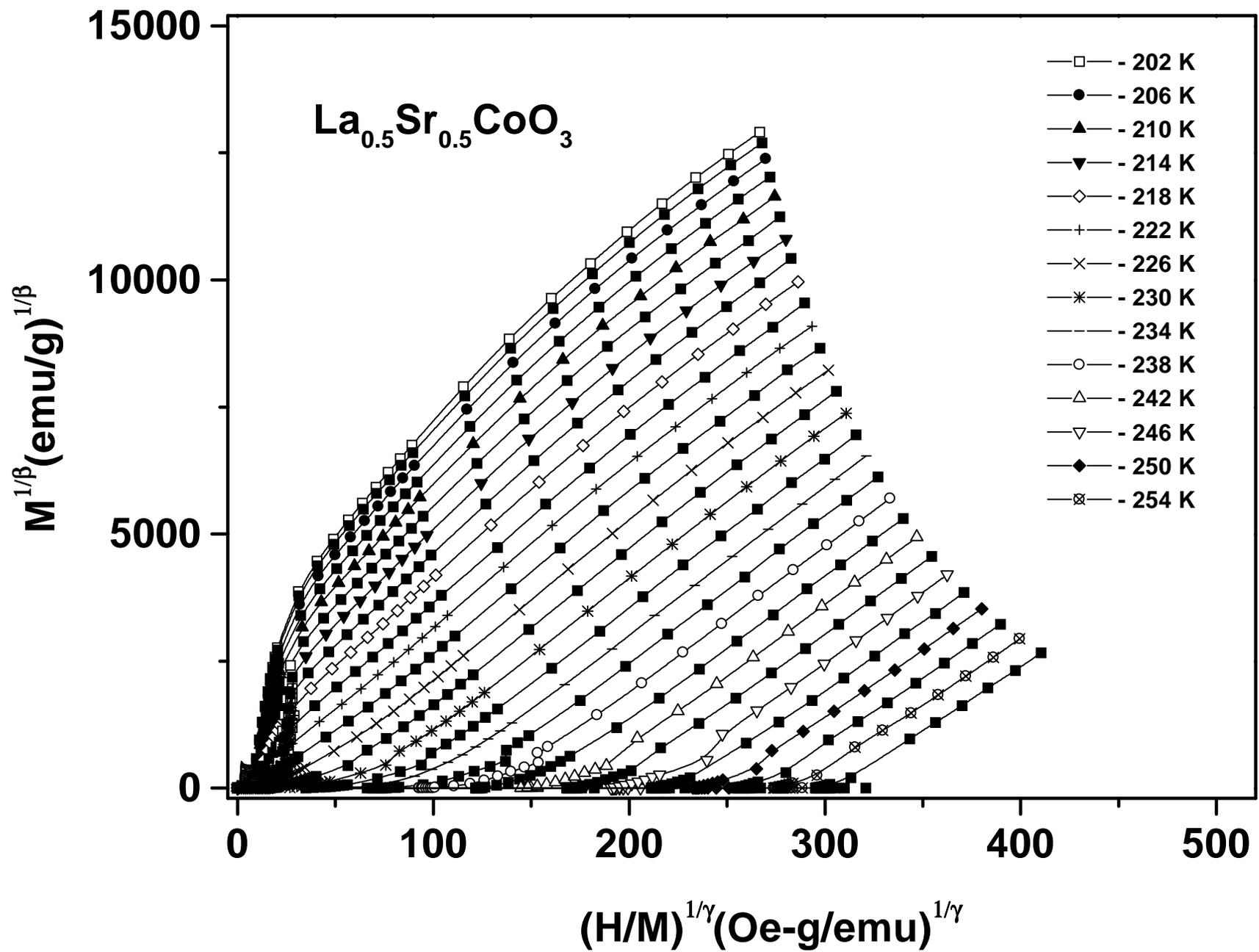
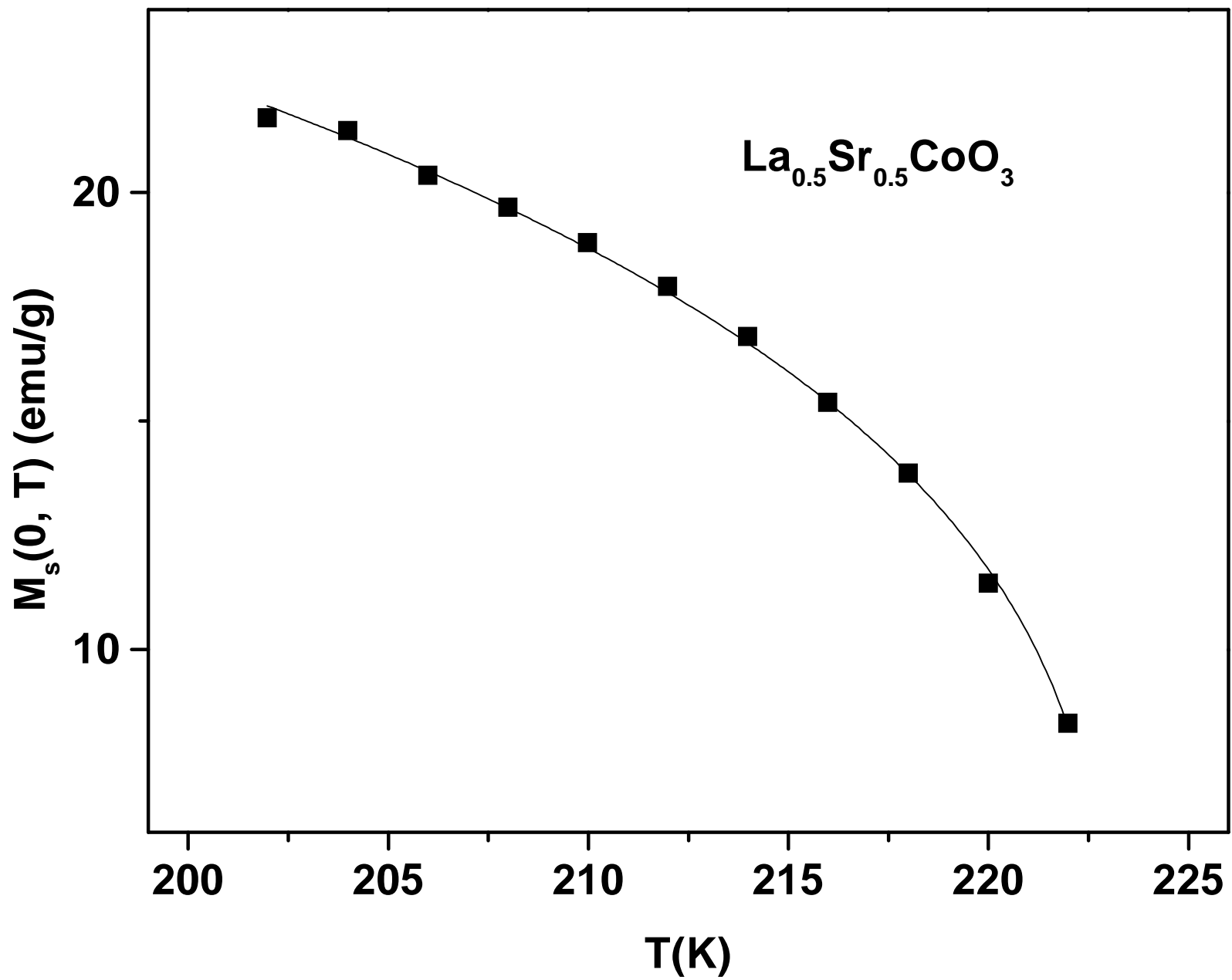
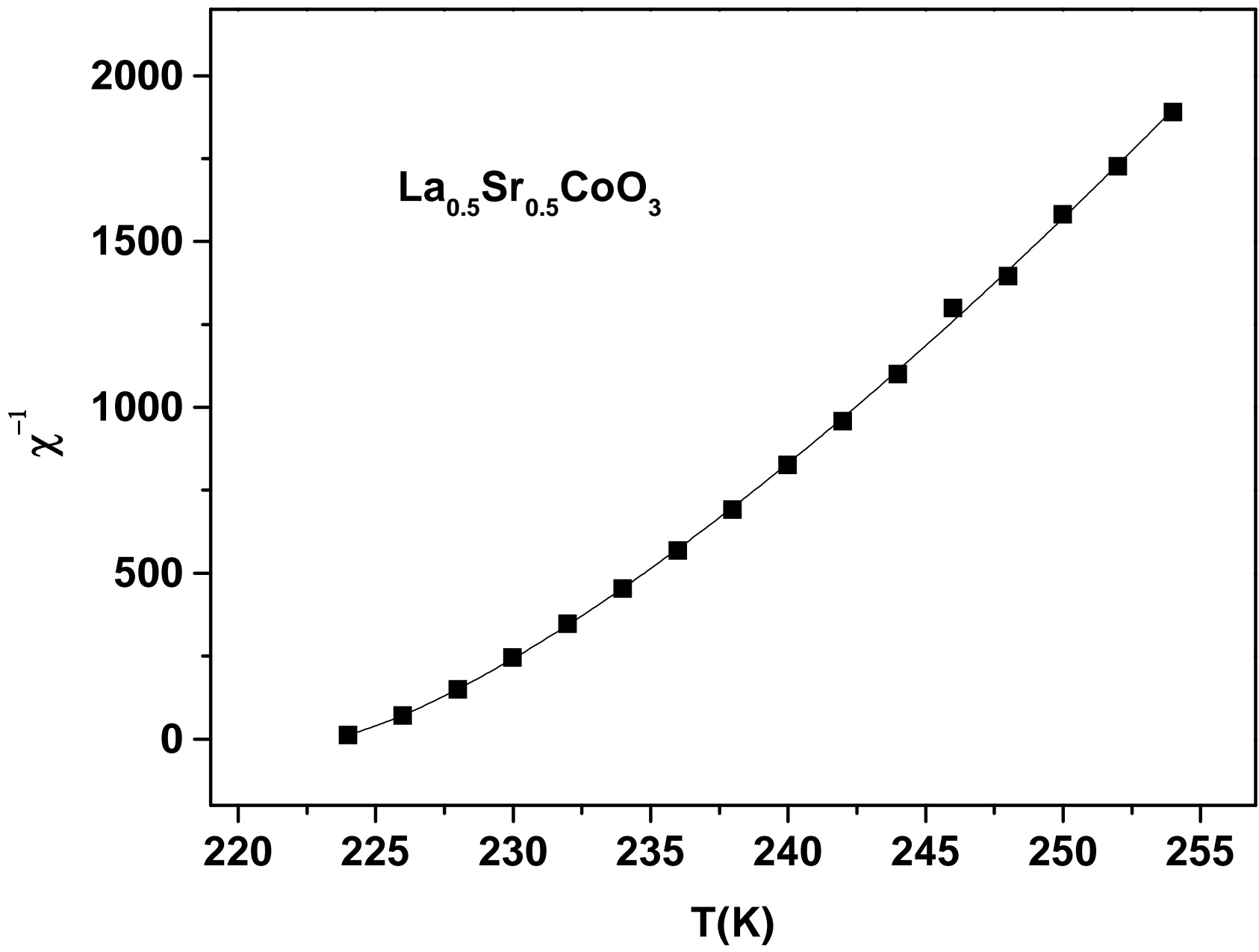
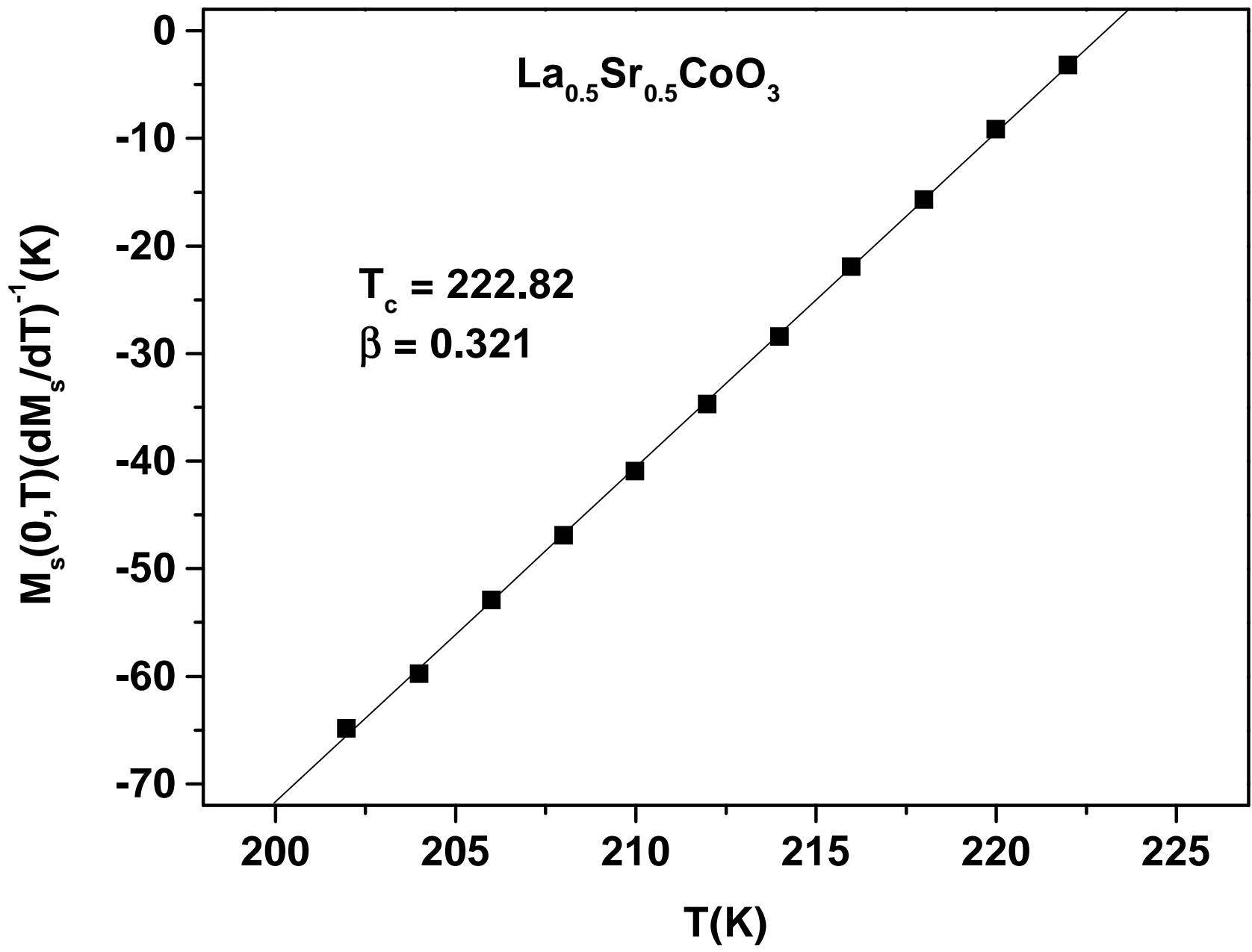
\end{document}